%% file: main.tex
\documentclass[]{spie}  %>>> use for US letter paper

\usepackage{amsmath,amsfonts,amssymb} \usepackage{graphicx}
\usepackage[colorlinks=true, allcolors=blue]{hyperref}
\usepackage{lineno}
\usepackage[loose]{units}
\usepackage[export]{adjustbox}
\usepackage{algorithm2e}
\usepackage{algorithmicx}
\usepackage{svg}
\usepackage{graphicx}
\usepackage{listings}
\usepackage{xcolor}
\usepackage{orcidlink}

\definecolor{codegreen}{rgb}{0,0.6,0}
\definecolor{codegray}{rgb}{0.5,0.5,0.5}
\definecolor{codepurple}{rgb}{0.58,0,0.82}
\definecolor{backcolour}{rgb}{0.95,0.95,0.92}

\lstdefinestyle{mystyle}{
    float=tp,
    floatplacement=tbp,
    backgroundcolor=\color{backcolour},
    commentstyle=\color{codegreen},
    keywordstyle=\color{magenta},
    numberstyle=\tiny\color{codegray},
    stringstyle=\color{codepurple},
    basicstyle=\ttfamily\footnotesize,
    breakatwhitespace=false,
    breaklines=true,
    captionpos=b,
    keepspaces=true,
    numbers=left,
    numbersep=5pt,
    showspaces=false,
    showstringspaces=false,
    showtabs=false,
    tabsize=2
}

\graphicspath{{./figs}}
\title{
  The Simons Observatory: A Minimum-Cost Matching Algorithm for Pairing Measured Resonances with Designed Detectors
}

\author[1]{Jack Lashner\,\orcidlink{0000-0002-0001-2151}}
\affil[1]{Wright Laboratory, Department of Physics, Yale University, New Haven, Connecticut 06511}

\author[2]{Kaiwen Zheng\,\orcidlink{0000-0003-4645-7084}}
\affil[2]{Joseph Henry Laboratories of Physics, Jadwin Hall, Princeton University, Princeton, NJ 08544}

\author[3]{Kevin T. Crowley\,\orcidlink{0000-0001-5068-1295}}
\affil[3]{Department of Physics, University of California, San Diego, La Jolla, CA}

\author[4, 5]{Nicholas Galitzki\,\orcidlink{0000-0001-7225-6679}}
\affil[4]{Department of Physics, University of Texas at Austin, Austin, TX, 78712, USA}
\affil[5]{Weinberg Institute for Theoretical Physics, Texas Center for Cosmology and Astroparticle Physics, Austin, TX 78712, USA}

\author[6, 7]{Kathleen Harrington\,\orcidlink{0000-0003-1248-9563}}
\affil[6]{Argonne National Laboratory, High Energy Physics Division. 9700 S Cass Ave, Lemont, IL 60439}
\affil[7]{University of Chicago, Department of Astronomy and Astrophysics. 5801 S Ellis Ave, Chicago, IL 60637}

\author[8]{Hironobu Nakata\,\orcidlink{0000-0002-6300-1495}}
\affil[8]{Department of Physics, Faculty of Science, Kyoto University, Kyoto}

\author[1]{Max Silva-Feaver\,\orcidlink{0000-0001-7480-4341}}

\authorinfo{Further author information: (Send correspondence to Jack Lashner)\\
Jack Lashner: E-mail: jack.lashner@yale.edu}

\begin{document}
\maketitle

\begin{abstract}
The Simons Observatory (SO) is a ground-based cosmic microwave background experiment currently being deployed to Cerro Toco in the Atacama Desert of Chile.
The initial deployment of SO, consisting of three 0.46m-diameter small-aperture telescopes and one 6m-primary large-aperture telescope, will field over 60,000 transition-edge sensors that will observe at frequencies between 30 GHz and 280 GHz.
SO will read out its detectors using Superconducting Quantum Interference Device (SQUID) microwave-frequency multiplexing ($\mu$mux), a form of frequency division multiplexing where an RF-SQUID couples each TES bolometer to a superconducting resonator tuned to a unique frequency.
Resonator frequencies are spaced roughly every 2 MHz between 4 and 6 GHz, allowing for multiplexing factors on the order of 1000.
One challenge of $\mu$mux is matching each tracked resonator with its corresponding physical detector.
Variations in resonator fabrication, and frequency shifts between cooldowns caused by trapped flux can cause the measured resonance frequencies to deviate significantly from their designed values.
In this study, we introduce a method for pairing measured and designed resonators by constructing a bipartite graph based on the two resonator sets, and assigning edge weights based on measured resonator and detector properties such as resonance frequency, detector pointing, and assigned bias lines.
Finding the minimum-cost matching for a given set of edge weights is a well-studied problem that can be solved very quickly, and this matching tells us the best assignment of measured resonators to designed detectors for our input parameters.
We will present results based on the first on-sky measurements from SAT1, the first SO MF small-aperture telescope.

%  SHORT ABSTRACT
% The Simons Observatory (SO) is a ground-based cosmic microwave background experiment currently being deployed to Cerro Toco in the Atacama Desert of Chile.
% SO will read out its detectors using Superconducting Quantum Interference Device (SQUID) microwave-frequency multiplexing ($\mu$mux), a form of frequency division multiplexing where an RF-SQUID couples each TES bolometer to a superconducting resonator tuned to a unique frequency.
% Variations in resonator fabrication, and frequency shifts between cooldowns caused by trapped flux can cause the measured resonance frequencies to deviate significantly from their designed values.
% We introduce a method for pairing measured and designed resonators by constructing a bipartite graph based on the two resonator sets, and assigning edge weights based on measured resonator and detector properties such as resonance frequency, detector pointing, and assigned bias lines.
% We will present results based on the first on-sky measurements from SAT1, the first SO MF small-aperture telescope.

\end{abstract}

\input{intro.tex}
\input{methodology.tex}
\input{results.tex}
\input{conclusion.tex}

\section{Acknowledgements}
This work was supported in part by a grant from the Simons Foundation (Award \#457687, B.K.).

\bibliography{main} % bibliography data in report.bib
\bibliographystyle{spiebib} % makes bibtex use spiebib.bst

\appendix

\input{solution_match_generation.tex}

% References
% \bibliography{report} % bibliography data in report.bib

\end{document}

%% file: intro.tex
\section{Introduction}

% - SO Overview
% - mumux Overview
% - Locating Resonators and basic readout
% - Importance of matching and matching summary

Millimeter wavelength radiation from the cosmic microwave background (CMB) contains a wealth of information about the early universe, and is key to understanding its structure and evolution.
The Simons Observatory (SO) is a collection of millimeter-wave telescopes, sited on Cerro Toco in the Atacama Desert of Chile, which will observe the temperature and polarization of the CMB in order to place constraints on cosmological parameters, and to map out the large-scale structure of the universe \cite{galitzkiSimonsObservatoryInstrument2018,adeSimonsObservatoryScience2019}.
Initially, SO will consist of three small-aperture telescopes\cite{galitzkiSimonsObservatoryDesign2024} and one large-aperture telescope\cite{zhuSimonsObservatoryLarge2021, gudmundssonSimonsObservatoryModeling2021}, which will be used to study the CMB from degree to arc-minute scales.
SO will use on order of 60,000 transition-edge sensing detectors (TES) to measure the incoming radiation in six different optical bands between 30 and 280~GHz.

To read out the 60,000 detectors, SO uses SQUID microwave multiplexing ($\mu$mux), a form of frequency-division multiplexing which achieves multiplexing factors on order of 1000 by inductively coupling each TES to a superconducting resonator with resonance frequency between 4 and 6\;GHz via an RF-SQUID \cite{doberMicrowaveSQUIDMultiplexer2017,hendersonHighlymultiplexedMicrowaveSQUID2018}.
Variations in the intensity of incoming radiation cause shifts in the TES resistance that are transduced into changes in resonance frequency. 
We use the SLAC Microresonator RF (SMURF) warm electronics\cite{yuSLACMicroresonatorRF} to monitor and interpret these changes in resonance frequency, to convert them into measurements of the TES current.

The focal planes for SO instruments consist of tiled, compact assemblies called universal focal-plane modules (UFMs) that contain the TES detectors, the optical coupling mechanism, and the cold readout electronics \cite{mccarrickSimonsObservatoryMicrowave2021,wangSimonsObservatoryFocalPlane2021}.
Each UFM contains two transmission lines, where each transmission line is coupled to 924 superconducting resonators tuned to frequencies between 4-6 GHz.
These transmission lines can be used to read out up to 1756 detectors, 1720 with optical coupling and 36 dark detectors.
The remaining uncoupled resonators are used for diagnostic purposes, such as to measure readout and two-level system noise.
The UFM also routes a number of DC bias lines, including two flux-ramp lines and 12 TES bias lines.

\begin{figure}[t]
    \centering
    \includegraphics[width=1\textwidth, trim={0 0 0 0}, clip]{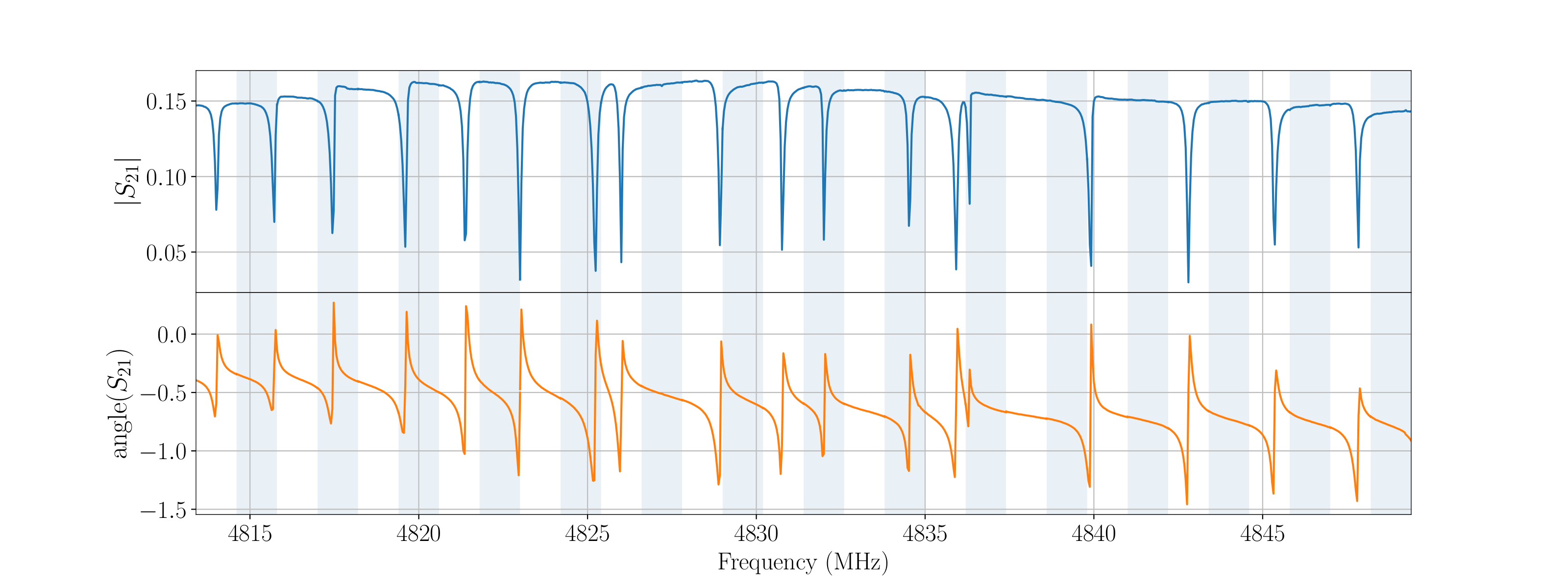}
    \caption{
        The complex transmission for a subsection of the 2 GHz band of the transmission line. Resonators show up as dips in the transmission with width around 30 kHz, and a designed spacing of 2 MHz.
    }
    \label{fig:find_freq}
\end{figure}

During operation, we need to tune the readout electronics to make sure it is outputting and tracking tones at the resonant frequencies of the superconducting resonators.
A \textit{tune}, or a measurement of the resonance frequencies, is created by measuring the complex transmission of a probe-tone as it sweeps across the 2 GHz band, and using a peak-finding algorithm to determine the location of each resonator.
The complex transmission for a small segment of the bandwidth can be seen in figure \ref{fig:find_freq}.

Calibration operations such as IVs are performed to determine resonator and detector characteristics, such as which bias line is connected to the TES, the normal resistance of the TES, and the operating bias voltages that should be used for each bias line.

A critical step in the analysis pipeline is determining what physical detector is connected to each tracked resonator.
This mapping is required to compare detector measurements across different tunes, and to combine data from multiple detectors into a single map of the sky.
The determination of this mapping is non-trivial, since even though we know the intended resonator frequencies of each detector, 
variations in the fabrication process and per-cooldown shifts in resonator frequency due to flux trapped in the wires and SQUIDs make it difficult to directly assign a measured resonance to a detector without additional information.
We must use a combination of resonator and detector properties, such as the resonance frequency, the coupled bias line, and the detector pointing information to determine the correct mapping from measured resonators to designed detectors.

In the following sections we will describe the algorithm used to generate the mapping between measured resonators and their physical detectors, and will present results based on initial observations showing its efficacy.

%% file: methodology.tex
\section{Methodology}
% - Problem formulation and generalization
% - Minimum cost matching algorithm with unassigned slots 
% - Cost function and unassigned cost function
% - pros and cons of matching
% - data preprocessing

In order to create maps, we need to be able to accurately determine which physical detector corresponds to each of the resonators that is being read out.
This determination is made based on a number of measured properties which can be compared to the corresponding designed properties for each of the detectors, including:
\begin{itemize}
    \item Resonance frequency of the superconducting resonator
    \item Detector pointing angle relative to boresight center
    \item TES bias line.
\end{itemize}

Previously, detector matches have been generated by hand based on lab measurements of resonance frequency and bias line assignments, which is challenging for a number of reasons.
Because lab measurements do not provide detector pointing information, these assignments are generated with considerably less information than is available during deployment.
Frequency matching without additional information can be time-consuming because the measured resonator frequencies scatter around the designed frequencies, resulting in clusters and crosses.
In some cases, due to resonator crossings and missing resonators, it may be impossible to determine the correct match without additional information.
This method of manual mapping is not feasible for deployment as it is time-consuming to generate for many UFMs, and we will need to regularly create new matches as more calibration data are obtained.
Furthermore, the inclusion of pointing information in deployment greatly reduces mapping uncertainty, enabling us to create an automated and robust software solution for detector mapping.

\begin{figure}[t]
    \centering
    \includegraphics[width=0.8\textwidth, trim={0 500 0 500}, clip]{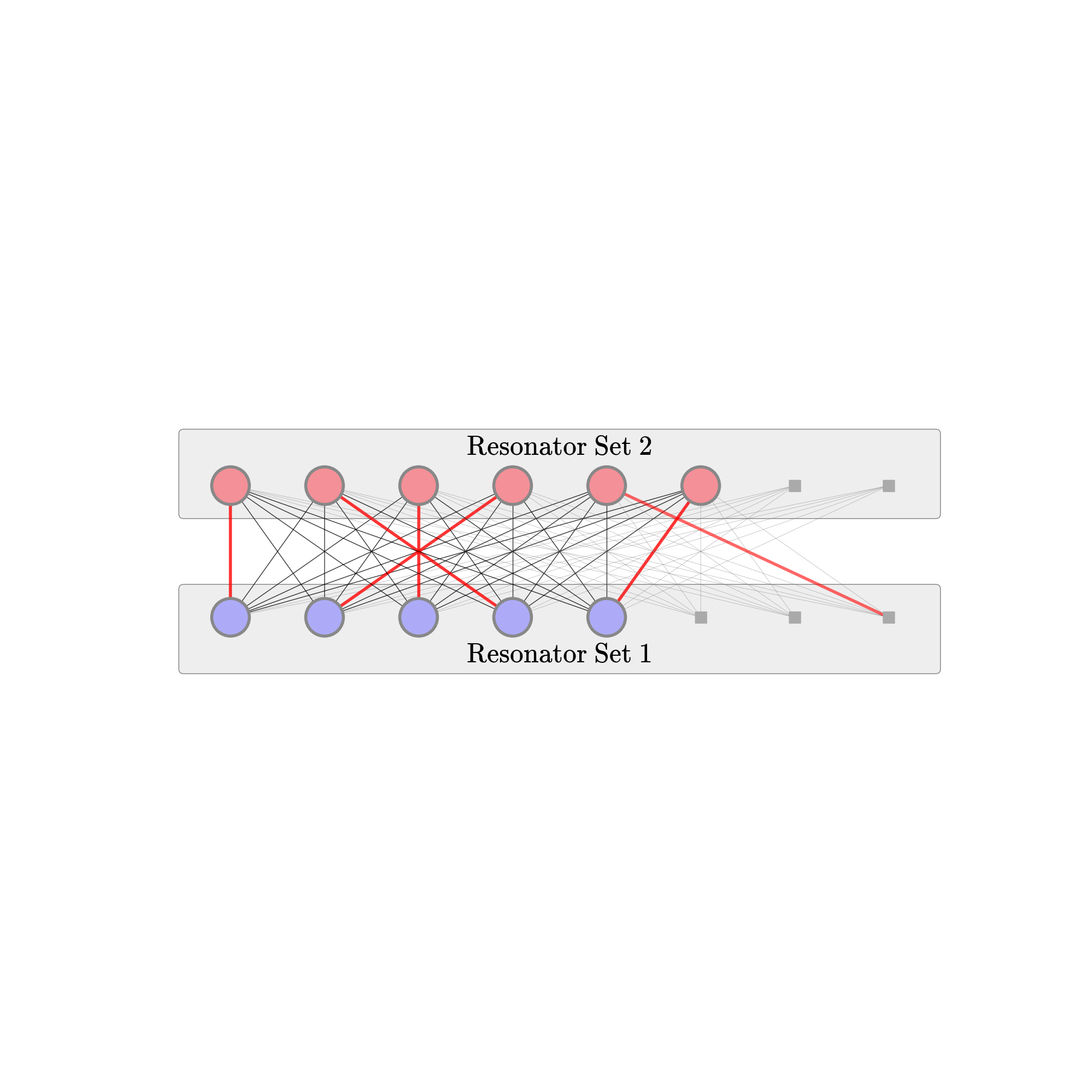}
    \caption{
        The complete graph representation of the matching problem.
        The two groups of nodes represent two sets of resonators to be matched, typically one being the set of measured resonators and the other being the set of designed detectors.
        Each colored node represents a different resonator and each grey square represents an unassigned node, used to represent when a resonator is not matched.
        Black and light grey lines show the resonator-to-node, and resonator-to-unassigned-node edges in the graph respectively.
        The red lines show an example matching between the two resonator sets.
    }
    \label{fig:matching_graph}
\end{figure}

To automate detector matching, we model the problem as a complete bipartite graph where each resonator is a node, and the disjoint sets represent the set of measured resonators and the set of designed ones.
Edge weights between the two sub-graphs are determined by a cost function that depends on resonator properties such as those listed above.
The best match is represented by the minimum-cost matching of the graph, the finding of which is a well-studied problem that can be solved very quickly for a given set of edge weights.
Due to the possibility that designed resonators are not present, or that spurious peaks are mistaken for real resonators in the measured data, we need a way to leave resonators on both sides unassigned.
To do this, we add a fixed number of \textit{unassigned nodes} to each side with edge weights that penalize leaving a resonator unassigned, where the penalty cost depends on the properties of the unassigned resonator.
A depiction of the bipartite graph and matching can be seen in figure \ref{fig:matching_graph}.
Our implementation can be found in \texttt{sotodlib}, the SO library for time-ordered data analysis\footnote{\url{https://github.com/simonsobs/sotodlib/blob/master/sotodlib/coords/det_match.py}}.

We generate our match based on edge weights using scipy's \texttt{linear\_sum\_assignment} function\footnote{\url{https://docs.scipy.org/doc/scipy/reference/generated/scipy.optimize.linear_sum_assignment.html}}, which implements a modified Jonker-Volgenant algorithm\cite{ramshawMinimumCostAssignmentsUnbalanced2012,crouseImplementing2DRectangular2016a}.
For a full UFM of around 1800 resonators per side, and an additional 1000 unassigned nodes per side, this algorithm takes roughly 0.5 seconds to find the optimal match.

The cost function that we use depends primarily on the three properties listed above: the difference in the resonance frequencies, the difference in the detector pointing angles, and whether the assigned bias lines based on calibration measurements matches the designed bias line.
Resonator frequencies are determined by sweeping a probe-tone from 4-6 GHz on each transmission line, measuring the complex transmission of the tone.
A peak-finding algorithm to determine the approximate location of each resonator, and then a finer sweep about each peak is used to hone in on the resonance frequency, and cut out resonators that don't have the expected shape.

Pointing measurements are made by scanning a celestial source, typically the Moon or Jupiter for the small-aperture telescopes, and finding the angle for which each detector sees the source relative to the boresight of the telescope.
The designed pointing angles are derived using an optical model of the instrument.
Detector pointing is parametrized by the angles $\xi$ and $\eta$: the angles of the detector beam center relative to the boresight in the elevation and azimuth directions.
They can be approximated as:
\begin{equation}
    \xi \approx \Delta \text{elevation} 
    \qquad 
    \eta \approx - \Delta \text{azimuth} / \cos(\text{elevation})
    \label{eq:pointing_angles}
\end{equation}

Bias lines are assigned to each measured detector by stepping the bias voltage up and down for each line one at a time while detectors are superconducting, and measuring which bias line has the largest correlation with TES current for each detector.

Below, we show the matching cost function and unassigned cost function that we use to determine the edge weights of the graph:
\begin{lstlisting}[
    language=Python,
    label={lst:matching_cost_funcs}
]

def matching_cost(r1, r2):
    if r1.coax_line != r2.coax_line:
        return np.inf

    cost = np.exp(np.abs(r1.freq - r2.freq) / delta_freq) 

    theta_diff = np.sqrt((r1.xi - r2.xi)**2 + (r1.eta - r2.eta)**2)
    cost += np.exp(theta_diff / delta_theta)

    if r1.bias_line != r2.bias_line:
        if r1.bias_line is None or r2.bias_line is None:
            cost += bl_unassigned_mismatch_penalty
        else:
            cost += bl_assigned_mismatch_penalty
    return cost

def unassigned_cost(r):
    if r.bias_line is None:
        return bl_unassigned_unmatched_penalty
    else:
        return bl_assigned_unmatched_penalty

\end{lstlisting}

Here the matching costs depend on the resonator properties \texttt{coax\_line}, the transmission line that the resonator is connected to; \texttt{freq}, the resonance frequency; \texttt{bias\_line}, the bias line providing DC voltage bias to the TES; and \texttt{xi} and \texttt{eta}, the pointing angles of the detector.
The parameters \texttt{delta\_freq}, \texttt{delta\_theta}, and the various bias line mismatch penalties are tunable parameters that can be adjusted to change the behavior and priorities of the matching algorithm.
We force a matching cost of $\infty$ for resonators that we know are on different transmission lines.
Different penalties are used for resonators where we were able to determine an assigned bias line, and resonators where we were not able to determine a bias-line, since it is more likely that measured resonators without proper bias-line assignments are spurious peaks rather than real resonators.

The matching algorithm is typically run in one of two modes of operation.
The first is to find the best possible solution between measured resonators of a UFM and designed detectors, called a \textit{solution match}.
This takes into account any calibration data available, including pointing measurements generated from scanning a celestial sources, bias line assignments.
The \textit{solution match} can be thought of as a fingerprint that is unique to each UFM: though most UFMs share the same designed resonator frequencies the fabrication variation is different for every UFM, so the solution match is unique.
For the solution match it is important to run preprocessing on the measured data to account for any global shifts and systematic offsets in both frequency and pointing space, as any systematic offset in the datasets can easily lead to incorrect matches.
The methods that we use for data preprocessing and the matching parameters used to generate the solution match are discussed in Appendix \ref{sec:data_preprocessing}.

The second mode of operation is to match a set of measured resonance frequencies to an existing solution match.
While the cooldown-to-cooldown variation in resonance frequency is much smaller, on the order of tens of kHz, this variation can still effect the SMuRF channel assignment in such a way that it makes detector comparisons impossible across different tunes without a matching.
To account for this, we generate a new match to the measurement set used to generate the solution match for each new tune, allowing us to determine the designed detector for each measured resonator in the new tune.
Unlike the solution match, this mode of operation matches two measured resonator sets with each other, and so there is only cooldown-to-cooldown variation and no fabrication variation.
This means we are able to rely just on resonance frequencies for this match, and don't need to incorporate other data sets.

New tunes are created once at the beginning of each cooldown, and occasionally during the cooldown if we observe higher noise measurements and suspect we can benefit from a new measurement of the resonance frequencies.
So far during the deployment, we have not found it necessary to regularly create new tunes during a cooldown, and tunes have remained stable for months at a time.

%% file: results.tex
\section{Results}

\begin{figure}[t]
    \centering
    \includegraphics[width=0.45\textwidth, trim={20 0 20 0}, clip]{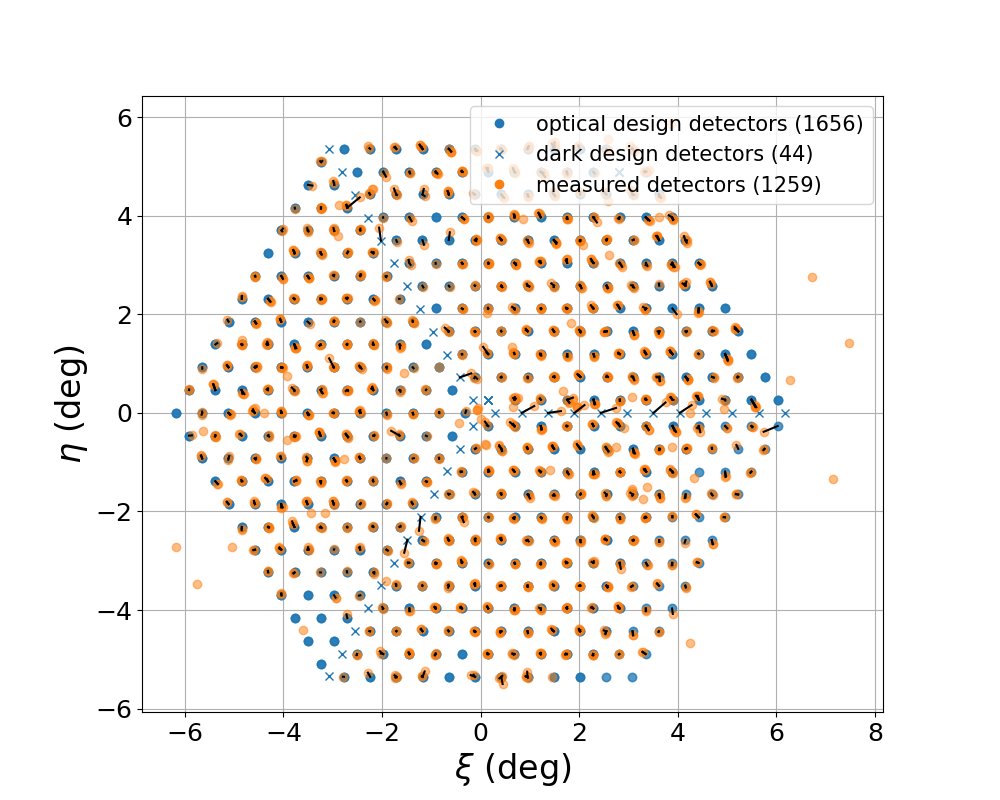}
    \includegraphics[width=0.45\textwidth, trim={20 0 80 0}, clip]{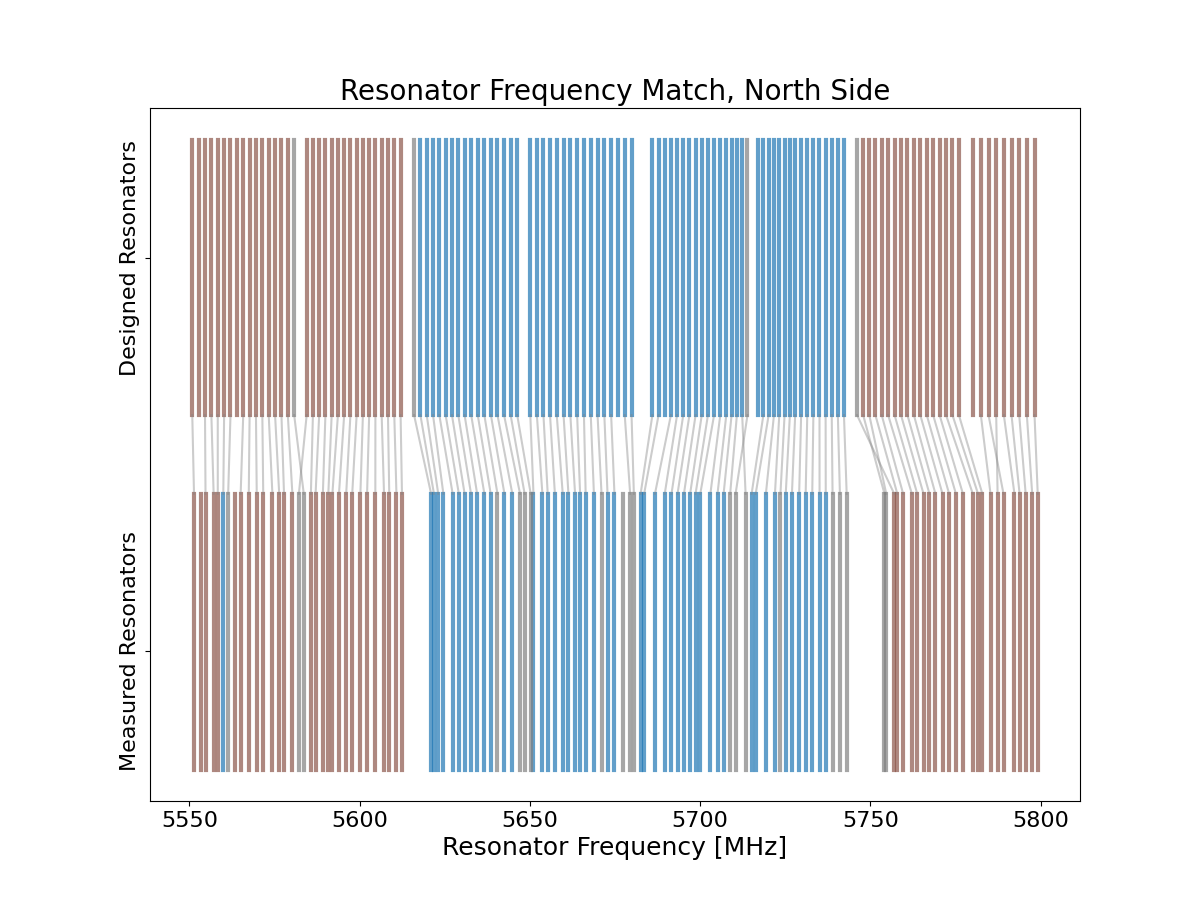}
    \caption{
        Matching results for the central UFM in SAT-MF1.
        (Left) Pointing angles of the designed detectors (blue) and measured detectors (orange), and the best matches between the two (black).
        Measured pointing angles are based on beam maps made by scanning the Moon.
        Dark design detectors are depicted with ``x'' markers.
        Due to the brightness of the Moon, the source can sometimes be seen in the dark-detector data due to its thermal effect.
        Note that these detector pointing angles have been preprocessed and aligned using the method described in Appendix \ref{sec:data_preprocessing}.
        (Right) Resonator frequencies of the designed detectors (top) and measured detectors (bottom), and the best matches between them (light-grey lines in the middle) for a small region of frequency space.
        Different colors of the resonator lines represent different assigned bias lines, where grey means that the resonator is not coupled to a bias line, either intentionally or due to wire-bonding issues. This region of frequency space contains only two separate bias lines in addition to resonators with no bias line assignments.
    }
    \label{fig:matching_results}
\end{figure}

Here we show results from a single, central UFM in the SAT-MF1 instrument, as this UFM has good detector yield and nearly complete pointing coverage.
To assess the performance of the matching algorithm and the accuracy of the solution matches, we look at various metrics, including the percentage of detectors which were matched, the percentage of resonators with matching bias line assignments, and the differences in the pointing angles between the measured and designed datasets.
A depiction of differences between the two detector sets in both pointing and frequency space can be seen in figure \ref{fig:matching_results}.
Statistics for these results can be seen in table \ref{tab:matching_statistics}.
The distribution of the differences in pointing angles between the measured and designed detectors can be seen in figure \ref{fig:ang_diff_hist}.

After performing the data preprocessing described in Appendix \ref{sec:data_preprocessing}, detector pointing angles align very well between the measured and designed datasets, as can be seen in the left panel of figure \ref{fig:matching_results}.
This allows us to match using a small tolerance in pointing angle difference, resulting in fewer mismatches.
Detectors that are left unassigned are those for which we were not able to make good pointing measurements, either due to detector data quality issues or because the detectors were out of range of our source scans.
We expect pointing angles for these detectors to be filled in as we take and analyze more calibration data, leading to better solutions.

Within a single pixel, we can be confident that we don't have cross-polarization or cross-band mis-matches, because resonators within a single pixel are designed to have very different resonance frequencies.
Additionally, any given bias line only controls detectors observing at a single frequency band, helping further avoid cross-band mismatches.
If we believe that we still see cross-band or cross-polarization mismatches, we may look into adding additional datasets to the matching algorithm, such as wiregrid polarization angle measurements, or detector time constants to break the degeneracy.

\begin{figure}[!t]
    \centering
    \includegraphics[width=0.8\textwidth]{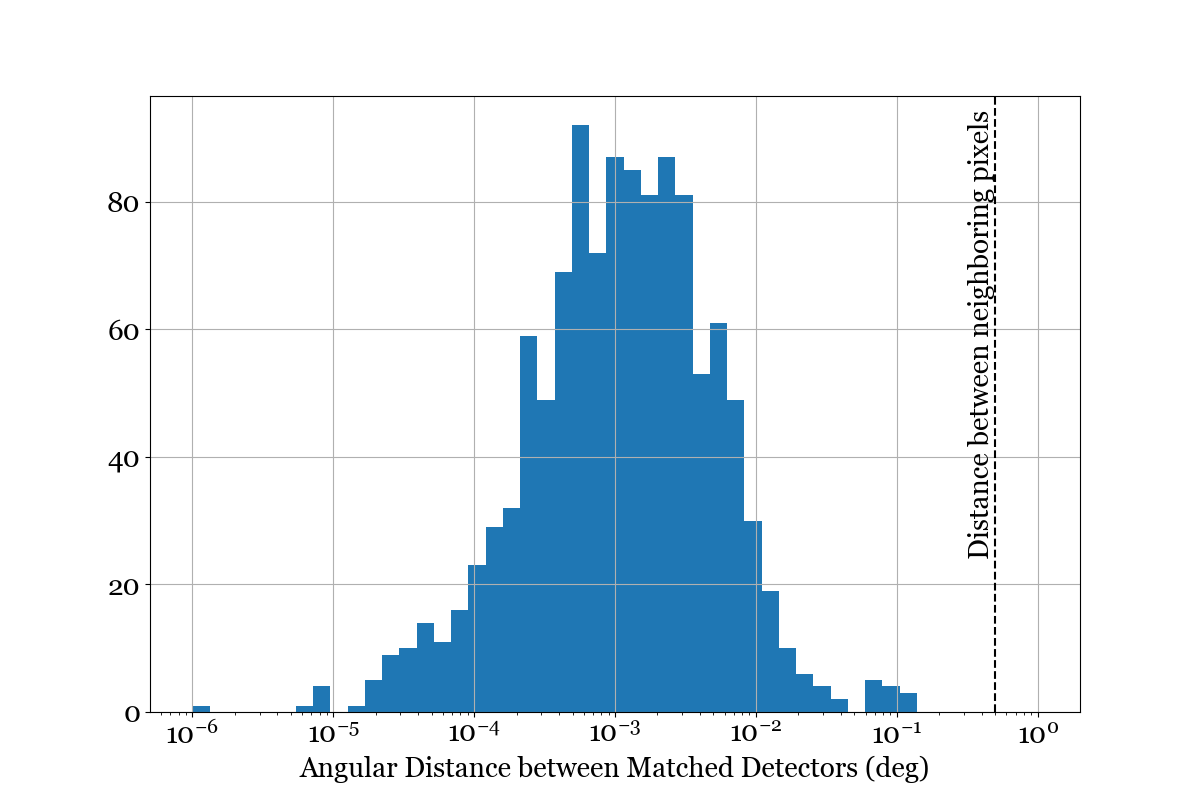}
    \caption{
        Distribution of angular separation between measured and designed detector pointing angles for the best match.
        Before matching, deformations in the measured pointing angles are corrected using the procedure described in Appendix \ref{sec:data_preprocessing}.
        Angle differences of all detector pairs fall well below ~0.5 degrees, which is the separation between neighboring pixels on the focal-plane.
        This signifies that there are few cross-pixel mismatches.
    } \label{fig:ang_diff_hist}
\end{figure}

\begin{table}[!htb]
    \centering
    \input{figs/results_table.tex}
    \vspace{1em}
    \caption{
        Table comparing the number of resonators in each set, the number of matched resonators, and the number of resonators which have pointing estimations that are left unmatched.
        Few detectors with good pointing information are unable to find matches.
        For most unmatched detectors with pointing information, even though we have pointing fits, these fits have high uncertainties and are not suitable for a match with narrow pointing tolerances.
        As we obtain better detector pointing data, we can expect the percentage of matched detectors to increase.
        The large difference between the number of measured and designed resonators is partially caused by two missing mux-chips on this particular UFM, accounting for 132 missing resonators.
    }
    \label{tab:matching_statistics}
\end{table}

\begin{figure}[!htb]
    \centering
    \includegraphics[width=0.8\textwidth]{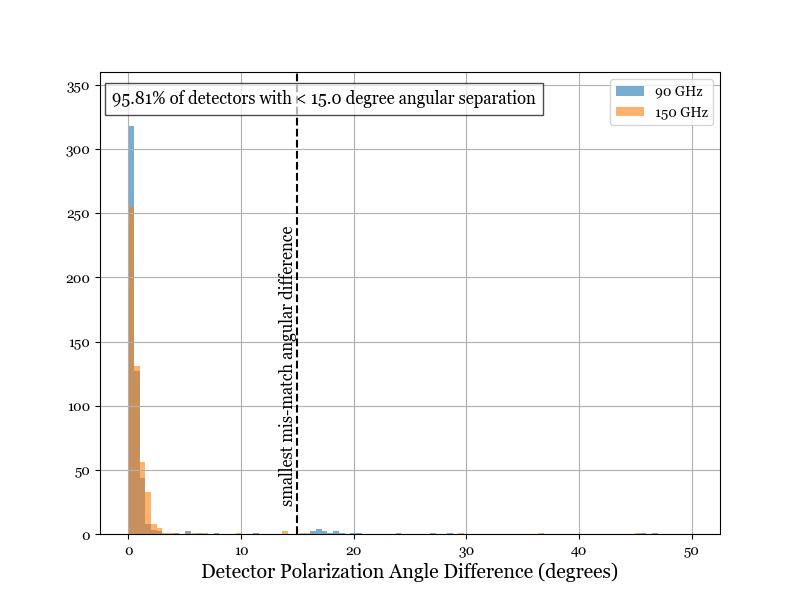}
    \caption{
        The distribution in the difference between the detector polarization angle measured with a wire grid.
        Here, we are taking only detectors with good pointing information, as those matches are the most reliable.
        We subtract out a global offset for each frequency band, to account for a difference in definition of the polarization angle, and to account for rotation that is induced by the detector time constant.
        We must subtract the medians for the frequency bands separately because the detector time constants are fairly different between the two frequency bands.
        Due to the arrangement of the focal-plane, a mismatch between detectors that have different polarization angles would appear at angles of $n \pi / 12$ radians, which is shown by the vertical dashed line for $n=1$.
        We can see a small cluster of mismatched 90 GHz resonators, however 95\% of detectors lie below the mismatch threshold.
    }
    \label{fig:wiregrid_angles}
\end{figure}

We can also validate detector matching results against datasets that were not involved in the match creation process. 
One such dataset is a measurement of our detector polarization angles measured using a rotating sparse wire grid that reflects linearly polarized light along the wire direction\cite{murataSimonsObservatoryFully2023}.
We compare the detector polarization angles measured using the wire grid calibration with the angles of the designed detectors on the focal-plane and see good agreement after subtracting a global angular offset from the measured data, as can be seen in figure \ref{fig:wiregrid_angles}.

%% file: figs/results_table.tex
\begin{tabular}{|l|c|}
\hline
Designed Resonators & 1783 \\\hline
Measured Resonators & 1495 \\\hline
Matched Resonators & 1427 \\\hline
% Unmatched with pointing & 63 \\\hline
\end{tabular}

%% file: conclusion.tex
\section{Conclusion}

SO will be analyzing data from 60,000 detectors to generate detailed maps of the CMB.
SO will be using SQUID microwave multiplexing to readout on order of 1000 detectors per transmission line, by coupling each TES to a superconducting resonator with a unique resonance frequency between 4 and 6 GHz.
A critical component of the analysis is determining which physical detector is mapped to each measured resonance frequency along the transmission line.
This is non-trivial due to differences in the design and true resonances caused by fabrication variation as well as cooldown-to-cooldown shifts in resonance frequency.

In this paper, we have presented our automated method for finding this mapping using properties such as resonance frequency, TES bias line, and detector pointing angles.
We show matching statistics for the central UFM of the small-aperture telescope.
We show that, given good detector pointing data, we are able to find detector matches for over 95\% of measured resonators.
We assess the quality of the match by comparing the angular difference in measured and designed detector pointing angles  and detector polarization angles for matched pairs.
Based on these datasets, we find that mismatches are rare for detectors with good pointing information.
We expect that as we obtain more calibration data, we will be able to generate correct match solutions for all UFMs.

%% file: solution_match_generation.tex
\section{Data Preprocessing}
\label{sec:data_preprocessing}

When generating a solution match, it is important to preprocess the data before running the matching algorithm, as systematic shifts in the resonance frequency or detector pointing datasets can easily lead to incorrect matches.
In addition to random variation in the resonance frequencies, fabrication error can cause entire mux-chips, containing hundreds of resonators clumped together in frequency space, to be offset by a constant frequency.
Resonance frequencies can also shift between cooldowns due to changes in the amount of flux trapped in the transmission line and SQUID loops.
For detector pointing, the measured pointing angles and designed pointing angles may be off by an affine transformation.
To correct for these transformations, we apply a sequence of iterative matches, freeing up different axes of parameter space in each iteration by adjusting the tunable parameters, and using the results to correct the measured data.
Once resonator frequency and pointing angles are correctly aligned, we can narrow tolerances in the cost function to produce a final solution match.

First, we apply a global shift to the measured detector pointing angles, to account for global systematic offsets. In most cases, the shift can be found by finding the difference between the centroid of the measured and designed pointing angle clusters.
However, in cases where significant portions of the focal-plane are missing measured pointing data, we may need to set this global offset by eye.

\begin{figure}[t]
    \centering
    \includegraphics[width=0.8\textwidth]{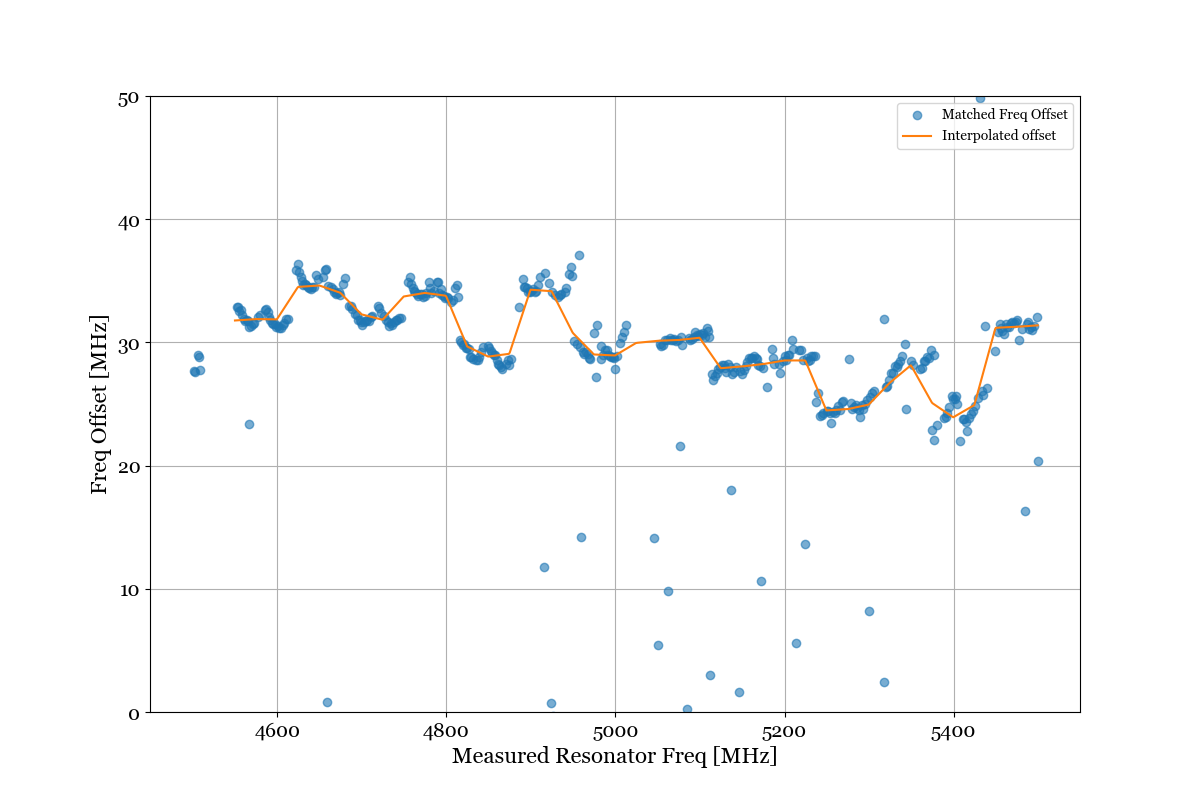} 
    \caption{
        Resonator frequency offset between measured and designed resonators for a region of frequency space before any corrections are applied.
        The orange line is the rolling median of the offset using a box-size of 100 MHz.
        This is what is subtracted by the first correction step to roughly align measured and design frequencies.
    }
    \label{fig:freq_offset_correction}
\end{figure}

Second, in order to correct for global and per-mux-chip frequency offsets, we greatly relax the resonance-frequency penalty and apply a rough match based on pointing angles and bias line assignments.
Here we tune $\Delta \theta$ in the listing \ref{lst:matching_cost_funcs} to be several degrees, much larger than the difference in angle between neighboring detectors, which is roughly 0.5 degrees.
We subtract out a rolling median of the frequency error between resonators and their match over large chunks of frequency space (we use a box size of 100 MHz) to correct for global offsets, as can be seen in figure \ref{fig:freq_offset_correction}.

\begin{figure}[t]
    \centering
    \includegraphics[width=0.8\textwidth]{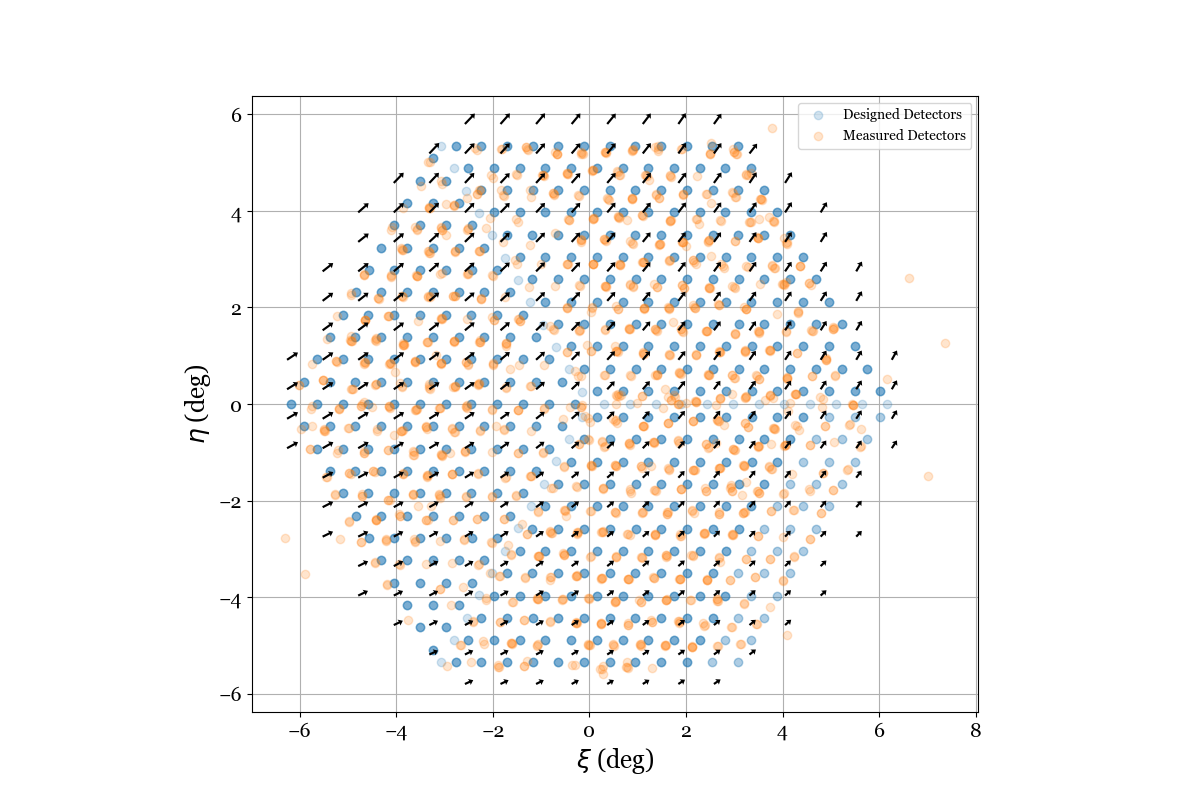}
    \caption{
        Detector pointing corrections between measured and designed detectors.
        The detector pointing angles shown here are those after an initial global correction that aligns the centers of the two distributions.
        The black arrows show the local interpolated pointing offsets for the pre-corrected match, computed by taking the median in the pointing angle difference between design and measured detectors about a circle centered at each point with a radius of 2 degrees.
        These are the final pointing corrections that are applied before the final match, shown in figure \ref{fig:matching_results}.
    }
    \label{fig:pointing_offset_correction}
\end{figure}

Next, we use a similar method to correct for large-scale pointing deformations.
We perform a match based primarily on resonance frequency and bias line assignment, relaxing the pointing restriction such that it is much larger than the difference in angle between neighboring detectors.
We then compute the median pointing error between a resonator and its match over large swaths of pointing space (we use a radius of 2 degrees), and subtract it out to correct for local pointing deformations.
The local deformations that are applied to the measured detector pointing before the final match can be seen in figure \ref{fig:pointing_offset_correction}.
This process corrects for residual pointing offsets that are not accounted for by the global shift, including any small scaling and rotational transformations.
Corrected resonator frequencies and detector angles can from this process can be seen in figure \ref{fig:matching_results}, where we see that the datasets are now well-aligned.

